\documentclass[12pt]{article}
\usepackage{epsfig}
\usepackage{amsmath}

\begin{document}
\begin{center}\Large {\bf On the history of geometrization of
    space-time:\\ From Minkowski to Finsler geometry. (100 years after
    Minkowski's Cologne adress.)} \large \\ Hubert 
Goenner\\ Institute for Theoretical Physics\\ University of
G\"ottingen\\ Germany 
\normalsize \end{center}

\section{Introduction: on the geometrization of physics}
This tribute to Hermann Minkowski will consist of three parts: a brief
historical introduction concerning geometrization of physics, a middle 
catering to mathematical themes, and a final chapter dealing with a 
(speculative) endeavour at applying Finsler geometry to physics loosly 
connected to Minkowski. 

From the history of physics we know that, at first, physical systems were
described in a given space and by a given time which both were regarded as {\em 
independent} of matter or any physical influence - not just by the philosopher
Kant. At his time, the idea of geometrizing space would have been
absurd. Johannes Kepler who for some period in his life had related
physical bodies, the planets, to geometric objects, i.e., to the five
regular polyhedra, certainly was far from what we now understand by 
geometrization of
physics, i. e. the embedding of physical objects (matter, fields) into
a geometrical framework. A weakening of the rigid understanding of space
seems to have occured when the  
notion of {\em non-euclidean} geometry came up, in the 19th century (C. F. 
Gauss, N. I. Loba$\hat{c}$evski, J. Bolyai). The answer to the question of
what kind of geometry the space we live in exhibits, now could be delegated to an 
empirical test \cite{Riemann}. Whether the anecdote about Gauss with his geodesic 
measurement of the angles of a triangle formed by three hills is true or not,
in any case the astronomer K. Schwarzschild investigated the question 
scientifically with bodies far away in the heavens (1900)
\cite{Schwarzschild00}. Also in the 19th century, the mechanics of rigid
bodies became reformulated within non-euclidean geometry (F. Klein, W. A. 
Clifford, R. S. Heath) \cite{Ziegler}). Yet, with the exception of Clifford, there still was no question about space or time being {\em influenced} by material 
systems.

As is well known, the joinder of space and time to space-time by Hermann
Minkowski whose famous speech about the ``union of space and time'' \cite{Minko1908} 
was commemorated in September 2008, became a first step in this still ongoing 
process of geometrization. As holds for many innovations in science,
the idea of space-time did not appear like a shooting star. A few 
mathematicians, fiction writers, and philosophers presented it quite
clearly before Minkowski, but not as a mathematical theory. In the 18th
century D'Alembert mentioned time as a fourth dimension in 1754 in his 
encyclopedia (written with Diderot); then again this was done by
Lagrange in 1797\footnote{``Ainsi on peut regarder la m\'ecanique
  comme une g\'eom\'etrie \`a quatre dimensions [..]''.} \cite{Archi1920}. For
the 19th century, let me first mention Charles Howard Hinton's article
of 1880: ``What is the fourth dimension?'' \cite{Hinton1884} which considered a fourth
{\em spatial} dimension. In his reply to it in the journal {\it Nature} of
1885, an anonymous letter writer signing ``S.'', introduced {\em time} as the 
fourth dimension and dealt with a 4-dimensional ``time-space''. S. mastered 
(verbally) what we now call the space-time picture, and even managed to 
correctly describe the hypercube by looking at the motion of a cube in 
time-space \cite{S1885}. In another article, Hinton tried to geometrize electrical
charge and currents \cite{Hinton1902}. Better known is H. G. Wells' novel ``Time
Machine'' of 1894 in which again a 4-dimensional junction of time and space 
called ``Space'' is considered \cite{Wells1895}. There, it is made clear that 
time is {\em not} considered as a fourth spacelike dimension. As the German
translation of Wells' book came out in 1904 \cite{Greve1904},
Minkowski could have read it, in principle. Finally, a philosopher of
Hungarian origin, Menyh\'ert (Melchior) Pal\'agyi, who had become a professor in 
Darmstadt published his ``New theory of space and time'' in 1901
\cite{Palag1901}. He joined space and time rather vaguely to a 4-dimensional 
entity named ``flowing space'' (flie\ss{}ender Raum), drew a Minkowski-diagram 
and introduced a ``time angle'' between the worldline of a moving particle and 
the time axis. He abstained from giving a mathematical scheme except for pointing
out that ``the coordinates of a point in flowing space could be represented by
$x + i~t, y + i~t, z + i~t$~'' (\cite{Palag1901}, p. 32).\footnote{This is a
  curious combination of time with one absolute and two relative
  spacelike coordinates. Let $x' = x + i~t~,~ y' = y + i~t~,~ z' = z +
  i~t~;$ then $y'-x' = y-x~; z'-x' = z-x$ are the relative
  coordinates. $x'-x = i~t$ could at best describe part of the light
  cone, a concept Pal\'agyi did not have. For the philosophy of
  Pal\'agyi cf. \cite{Boyce1928}.} After  he had become aware of
special relativity and, then, of Minkowski's famous speech, Pal\'agyi claimed 
priority for the space-time picture but rejected Minkowski's space-time 
manifold. From his writings it is obvious that his thoughts remain within 
psychology, and that he was incompetent both in mathematics and physics. 
\cite{Palag1914}. In sharp contrast, around 1905, and before Minkowski, 
Poincar\'e also had a 4-dimensional (space-time) formalism for the wave
equation and electrodynamics \cite{Poin05}.\footnote{In Germany, at the time, 
Poincar\'e's papers seem tohave been neglected. Many well known
scientists then (e.g., M. 
Planck, in his paper on relativistic mechanics) and even later historians of 
science do not refer to Poincar\'e's short paper of 1905 - before Einstein's -,
but only to Poincar\'e's extended presentation of 1906.}  Possibly, due to his
epistemological position as a conventionalist, he might not have been interested
at all in the issue of geometrization. 

\section{Minkowskian spaces}

Did Minkowski geometrize electrodynamics by formulating it on a 
space-time manifold? Not in the sense of having found a geometry in which the 
electromagnetic field corresponded to a geometric object. This would come only
 later - after the geometrization of the gravitational field - in the 
framework of unified field theory. Right after Minkowski, M. Planck
(1906) \cite{Goenner2008}, G. Herglotz (1910), F. Kottler (1912),
G. N. Lewis \& R. C. Tolman (1909) among others put mechanics and
electrodynamics into a space-time picture: they {\em relativized} such topics. 

The first geometrization in the narrower meaning was achieved by Einstein and
Gro\ss{}mann (1913-15) \cite{EinGross1914}. At first, Einstein had wanted to 
keep the
constancy of the velocity of light only for ``areas of almost constant 
gravitational potential'' (\cite{Einstein1915}, p. 713) Thus, in his attempt 
toward a relativistic theory of gravitation, he assumed the velocity of light 
to be a function of the (Newtonian) gravitational potential $\Phi$. He
replaced the space-time metric of Minkowski by \begin{equation}ds^2= c(\Phi)^2 
dt^2 - \delta_{\alpha \beta}dx^{\alpha} dx^{\beta}~(\alpha,\beta = 1,2,3)~, 
\label{EinsteinI} \end{equation} In an important next step, with Gro\ss{}mann's
help, he introduced a (semi-)Riemannian metric and identified its components 
with the now more numerous gravitational potentials.\footnote{For the
  most detailed and expert history of the formation of general
  relativity cf. the 4 volumes of \cite{Renn2007a},\cite{Renn2007b}.} 


\section{Minkowski space-time and Minkowski Space}

In this part, we first distinguish between the physicist's and the
mathematician's use of the expression ``Minkowski space'' and present some of
Minkowski's results concerning the geometry of normed spaces.

\subsection{Minkowski space-time}
\label{subsection:Minspacetime}
The introduction of an {\em imaginary} time-coordinate $T= i~ ct$ by Minkowski
 into the line element of space-time \begin{equation}ds^2= dT^2 +
  \delta_{\alpha \beta}dx^{\alpha} dx^{\beta}~ (\alpha,\beta = 1,2,3)
 \label{Minmetric} \end{equation} turned out to be a little misleading for
physics and more so for the general public. On the surface, (\ref{Minmetric})
looked as if physics now played in a 4-dimensional euclidean space - with four
spatial dimensions corresponding exactly to what Riemann had had in mind. The
 19th century had been full of talk and papers about 4-dimensional space with 
its striking  possibility to enter a locked (3-dimensional) room without 
breaking a seal \cite{Bragdon2005} \footnote{For a history of
  4-dimensional space and its modern uses cf. \cite{Robbin2006}}. It
was quickly realized, though, that a {\em real} representation of the
metric suited physics better, i.e., by a Lorentz-metric with signature 
$\pm 2$ (null cone, Cauchy problem etc.):
\begin{equation}ds^2= \eta_{ij}dx^i dx^j =
  c^2 dt^2 -  \delta_{\alpha \beta}dx^{\alpha} dx^{\beta} ~ (i, j =
 0,1,2,3)~.\end{equation} Space-time, as Minkowski had introduced it, became
 the framework for all physical theories in which velocities comparable to the 
velocity of light could occur: it is a natural representation space of the 
Lorentz (Poincar\'e-) group. Also in curved space-time it plays a role: as the 
tangent space at any point of the manifold of events. This is all too
well known such that nothing more needs to be said. 


\subsection{Minkowski Space}
\label{subsection:MinSpace}
The second meaning of the term ``Minkowski Space'' in the way mathematicians
 use it, is barely known to physicists. On this occasion of commemorating 
Minkowski in a tribute to him it is mandatory to include some of his 
mathematical achievements. 
\begin{quote} {\bf Minkowski 
Space is a real, finite-dimensional (d$\geq 2$)  normed (vector) space}
$V (= R^d)$ (\cite{Papad2005}, p. 138).\end{quote} 
If completeness is added, then it is just a special case of a {\em Banach}
 space. The norm $\parallel X \parallel$ of an element  $X \in~ V$
 satisfies the following conditions:\\ i) a)~~$\parallel X \parallel~ \geq 0;~~
 b)~\parallel X \parallel = 0$ if and only if $X=0$;~~\\ ii)~~ $\parallel \lambda X
 \parallel = |\lambda|\parallel X \parallel~;~ \lambda \in R ,~ X \in V$~;\\
 iii)~~$\parallel X+Y \parallel \leq ~  \parallel X \parallel+ \parallel Y
 \parallel$.\\ If 1) b) does not hold, a semi(pseudo)-norm of the kind needed
 in space-time obtains.

The unit ball $B \subset M^d$ with $ B:= \{ X\in V| \parallel X \parallel \leq 1\} $ is a (compact) {\em convex} and {\em symmetric} set
 \footnote{Minkowski called it ``Eichk\"orper''. A set $W$ is called symmetric
 (with regard to the zero point O) if $-W = W$. This means
 that all straight line segments passing through $O$ of the set are halved by
 $O$.} With the help of the norm, a (canonical) metric
 (semi-metric)\begin{equation} \delta (X,Y):=\parallel X-Y \parallel
 \end{equation} can always be introduced. The unit ball may be very different 
of what we imagine in an euclidean situation. 
In fact, if and only if the unit ball is an ellipsoid then Minkowski
Space turns out to be euclidean space (\cite{Busemann42}, p. 38). The
topology of any d-dimensional Minkowski Space is euclidean topology 
(\cite{Thompson96}, section 1.2). When Minkowski Space is seen as a
metrical space we speak of {\em Minkowski geometry}.  

Let me give an example for Minkowski geometry: Let $K$ be a compact, convex set in
euclidean space, $x\neq y$ two points of $K$ and $\xi \neq \eta$ points on 
the boundary $\partial K$ of $K$ met by straight lines joining $x$ and $y$
with the zero-point $O\neq x;~O \neq y$.
Then a {\em distance function} $F$ on
$K$ is defined by\begin{equation} F(x-y):= \frac{\parallel x-y
    \parallel}{\parallel \xi-\eta\parallel}~,~ F(0) = 0~.\end{equation} If the
euclidean norm is used, then \begin{equation} F(x-y) = 
\sqrt{\frac{\Sigma (x-y)^2}{\Sigma (\xi-\eta)^2}}~.\end{equation} The distance 
function is a convex function\footnote {A convex function satisfies the same 
conditions like a norm, i.e., the triangle inequality $F(x+y)\leq F(x)+ F(y)$, 
etc.} with $F(x)\leq 1$. 

The study of norms other than the euclidean (and not derived from a metric) is 
primarily due to Minkowski. Because of a 1-1 correspondence 
between norms on the linear space $V$ and symmetric, closed convex sets in $V$ 
with non-empty interior, convexity becomes an essential ingredient for the
 study of Minkowski Space\footnote{The correspondence reverses order:
 $B_1\subseteq B_2$ implies $\parallel \cdot \parallel_{B_1}\supseteq
 \parallel \cdot \parallel_{B_2}$ and $\parallel
 \cdot \parallel_{\alpha B} = \alpha^{-1}\parallel \cdot \parallel_{B}.$} 
\cite{Thompson96}. 
In this context, Minkowski introduced a vector sum of two convex 
sets (bodies) now called ``Minkowski sum''. Its combination with the concept
 of volume led him to important results. We realize that, with only a norm 
available, ``volume'' and
 ``orthogonality'' are not immediately at hand. As to volume, we assume that $V$
 is equipped with an auxiliary euclidean structure and that the volume is
 the Lebesque measure induced by this structure.\footnote{Closed convex sets
 being Borel sets, an alternative would be to use a Haar measure
 (\cite{Thompson96}, p. 53). - As to orthogonality in normed spaces, see the
 glossary for a definition.} In connection with his research in number theory, 
Minkowski used the concepts ``volume'' and ``area'' of convex bodies (cf. his 
``geometry of numbers'', \cite{Minkowski1910} and \cite{Minkowski1911}). To make progress, he introduced fundamental
 quantities as, e.g., the {\em support function} of convex bodies. Another one 
is {\em mixed volume} generalizing the euclidean volume: it comprises and 
connects the concepts of volume, area, and total mean curvature.\footnote{See
 glossary.} 

In this context, one of the well-known results of Minkowski among
mathematicians perhaps is the {\em Brunn-Minkowski} inequality concerning the
(n-dimensional) volumes $\lambda_n(K_0)$ and $\lambda_n(K_1)$ of 2 compact
convex sets $K_0,~K_1$ in n-dimensional euclidean space $E^n $.
Let $ 0\leq t \leq 1$, then:\begin{equation}[\lambda_n((1-t)K_0+tK_1)]^{\frac{1}{n}} \geq (1-t)[\lambda_n(K_0)]^{\frac{1}{n}} + t
  [\lambda_n(K_1)]^{\frac{1}{n}}~.\end{equation} Equality for some $0<t<1$
holds if and only if $K_0$ and $K_1$ lie in parallel hyperplanes or are
homothetic.\\ 

An example is given like follows: \small We place $K_0$ and $K_1$ in two parallel 
hyperplanes described by $x=0~,~x=1$, respectively, in (n+1)-dimensional 
euclidean space $E^{n+1}$. Then \begin{equation} (1-t) K_0 + t K_1 = conv(K_0
  \cup K_1) \cap \{x|x=t \}.\nonumber \end{equation} Let $D_t$ be the 
n-dimensional ball in  $E^{n+1}$ contained in the hyperplane $x=t$ centered on 
the $x$-axis, and with n-dimensional volume equal to that of $(1-t) K_0 + t
K_1$. Then the Brunn-Minkowski inequality states that the union of all $D
(0\leq t \leq 1)$ is a convex set. An illustration of this example is to be
found on page 416 of \cite{Gardner2006}.\normalsize\\

Although I am discussing Minkowski Space mainly because of its connection with
{\em Finsler spaces} to be introduced soon, I shall dwell on Minkowski's
geometrical insights a bit further. Minkowski introduced his geometry\footnote{The name ``Minkowski geometry'' was given only later by S. Mazur (\cite{Thompson96},
  p. 43).} when looking at a lattice formed by integers. An immediate
result is Minkowski's theorem about lattice points: 
\begin{quote} Let $\Gamma$ be a lattice in $R^d$,
  $K \in R^d$ a bounded, convex set symmetric with regard to the zero-point
  $O$ as its center. If its volume $\lambda_n(K)\geq 2^n \lambda_n(\Gamma)$, 
then $K$ contains at least one further lattice point (different from $O$).\end{quote} $\lambda_n(\Gamma)$ is the volume of the elementary cell;
in a cubic lattice with spacing 1 consequently $\lambda_n(\Gamma) = 1.$ An
immediate consequence is that the volume of a convex body the center of which
is a lattice point (and which is the only interior such one) cannot be
greater than $2^n$. Hilbert calls Minkowski's theorem ``one of the
most applicable theorems in arithmetics''\footnote{One of the
  applications is the approximation of real numbers by
  fractions.}. According to him, by using methods of Dirichlet,
Minkowski was able to conclude from this result that: ``[..] for 
an algebraic number field exists at least one prime number divisible
by the square of a prime ideal, a so-called branching number [..]'' 
(\cite{Hilbert1910}, p. 452-53). 

\section{Finsler Space}
Now we progress from Minkowski- to Finsler geometry, a geometry also
being used for a geometrization of physics. 
Physicists who want to learn something about Finsler geometry and look
into one or the other textbook known to them, very likely will find
themselves in a situation described by the mathematician H. Busemann almost 60
years ago, i.e., in: ``an impenetrable forest whose entire vegetation
consists of tensors'' (\cite{Busemann50}, p. 5). This is a result of
Riemann's mentioning of one type of such kind of space and the
ensuing almost exclusive application of methods used in (pseudo-) Riemannian
geometry. General Relativity has set the stage for theoretical physicists; 
hence up to now most of the research on Finsler space by relativists consisted 
in an extension of the space-time metric $g_{ij}(x^l)$ to metrics also
dependent on direction $g_{ij}(x^l, dx^m)$.  

The extension of Riemannian ``point''-space $\{x^i\}$ into a ``line-space'' 
$\{x^i, dx^i\}$ by L. Berwald and E. Cartan\footnote{Cartan called $\{dx^i\}$
a supporting element.}  did make things clearer but not easier: how do you
explain to a physicist a geometry supporting at least 3 curvature tensors and 
five torsion tensors? Not to speak of its usefulness for physics! Fortunately, the
``impenetrable forest'' by now has become a real, enjoyable park: through the 
application of the concepts of {\em fibre bundle} and {\em non-linear
  connection}. The different curvatures and torsion tensors result from
vertical and horizontal parts of geometric objects in the tangent bundle, or
in the Finsler bundle of the underlying manifold. This will be explained in 
\ref{subsubsection:KawaMatsu}. 

\subsection{Family lines}
Before three different approaches to Finsler geometry will be
discussed, the academic ancestry of both Paul Finsler (1894-1970) and
Herbert Busemann (1905-1994) is presented. They stand for two
fundamental approaches to Finsler geometry. Finsler was a doctoral
student of Theodor Carath\'eodory (1873-1950) who himself had obtained
his PhD with Hermann Minkowski, both in G\"ottingen. On the other
hand, David Hilbert (1862-1943) had Richard Courant (1888-1972) as one
of his doctoral students. With him Busemann wrote his Ph D. This again 
happened in G\"ottingen. Finsler started from the calculus of
variations; infinitesimal length, and the length of a curve are
fundamental concepts. Busemann followed are more axiomatic path by
widening the definition of distance.
  


\subsection{Finsler geometry}
In essence, Finsler geometry is analogous to Riemannian geometry: there, the 
tangent space in a point p is euclidean space; here, the tangent space is just a
normed space, i.e., Minkowski Space. Put differently: A Finsler metric for a 
differentiable manifold M is a map that assigns to each point $x\in M$ a norm
on the tangent space $T_xM$ 
(\cite{Papad2005}, p. 38). When I refered to the almost exclusive use of
methods from Riemannian geometry it meant that this norm is demanded to derive 
from the length of a smooth path $\gamma: [a,b]\rightarrow M$ defined by 
$\int_a^b \parallel \frac{d\gamma(t)}{dt}\parallel dt$. Then Finsler
space becomes an example for the class of length spaces
\cite{Papad2005}. 

\subsubsection{Following Finsler and Cartan}
\label{subsubsection:FinCar} 
In this spirit, P. 
Finsler \cite{Finsler1918} and E. Cartan \cite{Cartan1934} started from the
length of the curve \begin{equation} d_{\gamma}(p, q) := \int_p^q L (x(t),
  \frac{dx(t)}{dt}) dt~. \end{equation} The variational principle
$\delta d_{\gamma}(p,q)= 0$ leads to the
Euler-Lagrange equation \begin{equation} \frac{d}{dt}(\frac{\partial L}{\partial
    \dot{x}^i}) - \frac{\partial L}{\partial x^i} = 0~, \end{equation} which
may be rewritten into \begin{equation} \frac{d^2 x^i}{dt^2} + 2 G^i(x^l,
  \dot{x}^m)= 0\label{geodes} ~,\end{equation} with $G^i(x^l,\dot{x}^m) = \frac{1}{4} g^{kl}
(-\frac{\partial L}{\partial x^l} + \frac{\partial^2 L}{\partial x^l \partial
  \dot{x}^m } \dot{x}^m)~, $ and $ 2 g_{ik} =\frac{\partial^2 L}{\partial
  \dot{x}^l \partial \dot{x}^m }~,~ g^{il}g_{jl}=\delta_j^i~.$ The
theory then is developped from the ``Lagrangian'' defined in this
way. This includes an important object $ N^i_{~l}:= \frac{\partial
  G^i}{\partial y^l}$, the geometric meaning of which as a non-linear
connection we shall recognize in section \ref{subsubsection:KawaMatsu}.

In general, a {\em Finsler structure} $L(x,y)$ with $y:=\frac{dx(t))}{dt}=
\dot{x}$ and homogeneous of degree 1 in $y$ is introduced, from which the 
Finsler metric follows as: \begin{equation} f_{ij}=f_{ji} = 
\frac{\partial(\frac{1}{2}
    L^2)}{\partial y^i \partial y^j}~,~~~ f_{ij}y^iy^j = L^2~,~y^l
  \frac{\partial L}{\partial y^l}= L~,~ f_{ij}y^j = L \frac{\partial
    L}{\partial y^i}~.\label{Fimet}\end{equation}
A further totally symmetric tensor $C_{ijk}$ ensues: 
\begin{equation}C_{ijk}:=  \frac{\partial(\frac{1}{2}
    L^2)}{\partial y^i \partial y^j \partial y^k}~, \end{equation}
which will be interpreted as a torsion tensor. As an example for a
Finsler metric related to physics is the {\em Randers} metric:
\begin{equation} \large L(x.y) = b_i(x) y^i +
  \sqrt{a_{ij}(x)y^iy^j}.\label{Randers} \end{equation} 
 The Finsler metric metric following from (\ref{Randers}) is: \begin{equation}
   f_{ik} = b_i b_k + + a_{ik} +2 b_{(i}a_{k)l}\hat{y}^l-
   a_{il}\hat{y}^l a_{km}\hat{y}^m (b_n\hat{y}^n) \end{equation} with $\hat{y}^k:= y^k
 (a_{lm}(x)y^ly^m)^{-\frac{1}{2}}~.~ $\normalsize Setting
 $a_{ij}=\eta_{ij}~,~y^k=\dot{x}^k~,$ and identifying $b_i$ with the
 electromagnetic 4-potential $e A_i$ leads back to the Lagrangian for the motion
 of a charged particle.

In this context, a Finsler space thus is called a {\em locally Minkowskian space}
if there exists a coordinate system, in which the Finsler structure is a
function of $y^i$ alone. The use of the ``element of support'' $(x^i, dx^k
  \equiv y^k)$ by Cartan essentially amounts to a step towards working in the 
tangent bundle $TM$ of the manifold $M$.
\subsubsection{Following Minkowski and Busemann}
\label{subsubsection:MinBus}
In order to define a norm in Minkowski Space, H. Busemann did replace
the homogeneity condition by the relation: \begin{equation}
  \frac{\parallel PQ \parallel}{\parallel PQ' \parallel} =
  \frac{|PQ|}{|PQ'|} ~, \end{equation} where $P, Q$ and $Q'$ are
points on a line, $\parallel PQ \parallel$ is the Minkowski distance
between $P$ and $Q$, while $|PQ|$ measures the euclidean distance. An
example in two dimensions with coordinates $x,y$ is given by:
\begin{equation}\parallel PQ \parallel = \Phi[\frac{\nu_1 (x_1-x_2) +
    \nu_2 (y_1-y_2)}{\sqrt{(x_1-x_2)^2+
      (y_1-y_2)^2}}]~|PQ|~ \end{equation} with an arbitrary function $\Phi$
      and constants $\nu_1,~\nu_2$. 

Thus, more generally, from Busemann's point of view, ``the Minkowskian
distance originates from the
euclidean distance [..] by multiplying it with a factor which depends only on
the direction of the segment from $x$~ to $y$'' This means that the
Minkowski distance reads as \begin{equation} F(x-y) = F(u)~
  |y-x|~,\end{equation} 
where $u$ is a unit vector in the direction of $y-x$ (\cite{Busemann50}, p. 9). 
By transporting Busemann's idea to space-time, we arrive at a (pseudo-)
``Minkowski''-metric  \begin{equation} ds^2:=
  \Phi[\frac{a_{1l}dx^l}{\sqrt{\eta_{ab}dx^a dx^b}},~
  \frac{a_{2m}dx^m}{\sqrt{\eta_{ab}dx^a
      dx^b}},~....]~\eta_{ij}dx^idx^j \end{equation} 
with vectors $a_{1l},~a_{2m}$ (with constant components in a
particular coordinate system) and the Minkowskian flat space-time
metric $\eta_{ik}~.$ We will see in section \ref{subsection:Finslerian
special relativity} that Bogoslovsky's Finsler metric is a subcase of
this class.

It is here that the routes of researchers applying the methods of 
\ref{subsubsection:FinCar} and of Minkowski and Busemann separate. Following
Finsler and Cartan - as most of the relativists interested in Finsler geometry
have done -, we would use $ds^2 \simeq \frac{1}{2}~L^2$ as a Finsler
structure and {\em derive} the metric from it according to the first equation
in (\ref{Fimet}). 
\subsubsection{Following Kawaguchi and Matsumoto}
\label{subsubsection:KawaMatsu}
Here, a Finsler connection is defined as a pair $(N, \Gamma)$ of a non-linear
connection $N$ in $T(M)(TM, \pi_T, V^n)$ and a connection $\Gamma$ in
the Finsler bundle $F(M)(TM, \pi_1,\linebreak GL(n,R))$ linked to the tangent
bundle $T(M)$. Here, $\pi_1,~\pi_T $ are the projection maps from
$F(M)$ to $T(M)$, and from $T(M)$ to $M$, respectively.
The bundle of linear frames comes in
as soon as the directional elements are no longer restricted to the $dx^i$ in
$M$ (Cartan's ``supporting elements'') but are considered as arbitrary
vectors $y^i$ in some vector space. $V^n=R^n$ is the fibre of T(M)
over the manifold $M$. The projection maps from $F(M)$ to the bundle of linear
frames $L(M)$ and from $L(M)$ to $M$ are named $\pi_2$, and $\pi_L$. The
following relationship is demanded: 
\begin{equation} \pi_T \cdot \pi_1 =  \pi_L \cdot \pi_2~.  \end{equation} 

From the tangent bundle $T(M)$ and the decomposition of the Finsler bundle
$F(M)$ into a horizontal and a vertical subspace, the construction of 
{\em three} distinct connections is always guaranteed: $F\Gamma = (N^i_{~k}~,~
F_{jk}^{~~i}~,~C_{jk}^{~~i})$. With the three connections, three different curvature tensors
 $R_{ijk}^{~~l},~ P_{ijk}^{~~l},~ S_{ijk}^{~~l} $ and 8 torsion tensors, three
 of which vanish, may be formed.

With $F_{jk}^{~~i}$ and $C_{jk}^{~~i}$,
respectively, the horizontal and vertical covariant derivatives can be defined
\cite{Matsu86},~ \cite{Matsu03},~ \cite{Hashi1986}:

\noindent h(orizontal)-covariant derivative: \\
\begin{equation} \nabla_{\delta/\delta x^i} \left(\partial/\partial y^i\right) =
F_{ij}^{~k} \frac{\partial}{\partial y^k}  \end{equation} v(ertical)-covariant derivative:  \begin{equation} \nabla_{\partial/\partial y^i} \left(
\partial/\partial y^i \right) = C_{ij}^{~k} \frac{\partial}{\partial y^k} \end{equation} 

Now we can find the link to the Finsler-Cartan approach of
\ref{subsubsection:FinCar}. The {\em Cartan connection} is defined as $FC =
(N^i_{~k}~,~F^{*~~i}_{~jk}~,~C^{*~~i}_{~jk})$ with
\begin{eqnarray}F^{*~~i}_{~jk}:= \frac{1}{2}f^{il} (\delta_jf_{kl} +
  \delta_kf_{jl} - \delta_lf_{jk})~,~\delta_l := \frac{\partial}{\partial x_l}
  - N^i_{~k}\frac{\partial}{\partial y_l}~,\\ C^{*~~i}_{~jk} :=
  \frac{1}{2}f^{il}\frac{\partial_k f_{jl}}{\partial y_k}~.
 \end{eqnarray} 
It can be shown that the 
Cartan-connection is metric compatible:  \begin{equation} f_{ij \parallel k} =
  0~,~f_{ij |k} = 0~. \end{equation} Here, the first covariant derivative
  (``~$ \parallel~''$) corresponds to the v-covariant derivative, the second
  (``~$ |~''$) to the h-covariant derivative.\footnote{Many other connections 
have been defined like those named after Berwald, or
  Chern. Cf. \cite{Hashi1986} \cite{Mo2006}.} 

\section{Application of Finsler Geometry to physics}

In the last part of this talk, an application to relativistic physics
is presented, i.e., a possible break of Lorentz invariance modeled by 
``Finslerian relativity''

\subsection{Generalities}
Finsler geometry has been applied to many areas in classical physics and also
to biological systems. After looking at many of such papers I get the
impression that, up to now, in physics, this geometry was applied to 
systems with some sort of anisotropy (matter, fields) as an {\em
 auxiliary device} supposed to lead to a better
understanding.\footnote{Two examples from fluid mechanics \cite{Ciesz01}
  and material science \cite{MikChic02} may show this.} This is far
away from the use of Finsler geometry in a geometrization of
physics. For biological systems described by certain sets of ordinary 
differential equations, these
equations have been brought into the form of (\ref{geodes}), and then 
interpreted within Finsler geometry. In my view, in both cases, no new
insights into the physics or mathematics of the systems described has been 
reached which could not have reached without Finsler geometry. 
Perhaps, recent speculations about a possible break of Lorentz-invariance make
a difference. In connection with Finsler geometry, the key idea due to
G. Yu. Bogoslovsky is more than 20 years old\footnote{For a more
  recent presentation cf. \cite{Bogo1994}.}
\cite{Bogo1973},~\cite{Bogo1977}. Its importance is just about to be discovered 
by mainstream physics \cite{CohGlas06},~\cite{Gibbons07}.

\subsection{A possible break of Lorentz invariance?}
At first, the possible break of Lorentz invariance was motivated by the
presumed Greisen-Zatzepin-Kuz'min cut off in the energy of the observed
particle spectrum resulting from inelastic scattering of photons at 
ultra-high-energy cosmic-ray protons (production of Pions) calculated
from special relativity \cite{GKZ1966}. 
Although this has not yet been fully cleared up, the recent
measurements from the AUGER collaboration (detector array in
Argentina) seem to indicate that the
UHE-particles can be linked to Active Galactic Nuclei (AGN) thought to
be powered by supermassive black holes. This would weaken the
applicability of the GZK-cut off \cite{Auger07}. Other tests, e.g., using the
polarization of cosmic background radiation have been
suggested\footnote{Such data also are used to test a possible
  CPT-violation. Up to now, the data are not good enough to resolve
  the question.} \cite{KostMew07}.

From the point of view of theory, ad-hoc changes in the dispersion
relation for high-energy particles implying the break of
Lorentz-invariance have been discussed \cite{ColeGlash98},
\cite{CohGlash2006} as well as the inclusion of
direction dependent ``background fields'' into the quantum field theory
vaccum \cite{BluKostel2005}.
   
\subsection{Finslerian special relativity}
\label{subsection:Finslerian special relativity}
The basic idea of G. Bogoslovsky was to study a metric having as an isometry
group the largest subgroup of the Poincar\'e-group, an 8-parameter Lie
group (4-parameter subgroup of Lorentz group). It is now given the fancy
notation ISIM(2). In (1+1)-dimensions, the line element of homogeneity degree 
1 turned out to be: \begin{equation} ds =
  (\frac{dx^0 - dx}{dx^0 + dx})^{\frac{r}{2}}~{\sqrt{(dx^0)^2
        -(dx)^2}}~,\label{Bomet2}\end{equation} with $ 0 \leq r < 1$.
The velocity addition law of special relativity
remains unaltered. It is easy to extend (\ref{Bomet2}) to (1+3)-dimensions and
to write it manifestly covariant: \begin{equation} ds =  (\frac{a_l
    dx^l}{\sqrt{\eta_{rs} dx^r dx^s}})^r~\sqrt{\eta_{nm} dx^n
    dx^m}.\label{Bomet4} \end{equation} 
In (\ref{Bomet4}), $\eta_{lm}a_l a^m =
0$, i.e., $a^l = (1, {\bf a})~,~ {\bf a} \cdot {\bf a}=1$ is a null direction
in Minkowski space-time. 

The ``generalized Lorentz transformations'' now are 
\begin{eqnarray}
 x^{\prime^i} = x^i + t_0^i~,\\x^{\prime i} = D(\vec v, \vec
 a)~~R^i_{~j} (\vec v, \vec a) \underbrace{L^j_k (\vec v)}_{\tiny{\text{Lorentz boost}}}x^k = {\cal L}
(\vec v, \vec a) x^k   \label{SIMS}~,\end{eqnarray} with\\

\begin{tabular}{ll}
$R^i_{~j}(\vec v, \vec a)$:& {rotation of space axes about $\vec v
  \times \vec a$}\\ & {through an angle
determined by $\vec v$ and $\vec a$.} \\ $\vec v$: & {velocity of moving frame}
\end{tabular}\\ 

From the equation for the mass shell, a highly non-linear ``modified''
dispersion relation as compared to $\eta_{ij}p^i p^j= m^2c^2$
follows:
\begin{equation}[\frac{\eta_{ij}p^i p^j}{(\eta_{lm}p^la^m)^2}]^r \eta_{st}p^s p^t = m^2c^2
(1+r)^{1+r} (1-r)^{1-r}~. \label{moddisp}\end{equation} As the figure
shows, for growing $r$ the mass shell becomess more and more
anisotropic. In some of the papers on the breaking of
Lorentz symmetry, polynomial additions were suggested: $\eta_{ij}p^i
p^j + a_{ijk}p^ip^jp^k + b_2(\eta_{ij}p^ip^j)^2 + ... = m^2c^2$.    

The kinematics of this Finslerian special relativity disgresses from what we
are used to. The expressions for energy and linear momentum of a relativistic
particle now are given by
\begin{eqnarray}~E = \vec p \cdot \vec v
  -L~~,~~ \vec p = \frac{\partial L}{\partial \vec v}\\ 
E = \frac{mc^2}{\sqrt{1- \vec v^2/c^2}} \left( 
\frac{1- \vec a \cdot \vec v/c}{\sqrt{1- \vec v^2/c^2}}
\right)^r \left[1 - r + r \frac{1- \vec v^2/c^2}{1- \vec a \cdot \vec v/c}
\right]\\ \vec p = \frac{mc}{\sqrt{1- \vec v^2/c^2}} \left(
\frac{1- \vec v \cdot \vec a/c}{\sqrt{1- \vec v^2/c^2}}
\right)^r
\left[(1-r) \vec v/c + r \vec a \frac{1- \vec v^2/c^2}{1- \vec v \cdot
  \vec a/c}
\right] . \end{eqnarray} In the non-relativistic limit, we
obtain \begin{eqnarray} E = mc^2 + (1-r)\frac{m \vec v^2}{2} + r(1-r) 
\frac{m(\vec v \cdot \vec a)^2}{2} + O\left(\left(\frac{\vec
      v}{c}\right)^3\right)\\ \vec p = rmc \vec a + (1-r) m \vec v +
r(1-r) m (\vec v \cdot \vec a)\vec a +O\left(\left(\frac{\vec
      v}{c}\right)^2\right) \end{eqnarray} With a non-vanishing 
anisotropy-parameter $r$, there exists a ``rest-momentum'' even for
vanishing velocity.\footnote{In fact, the theory is built for a {\em
    relativistic} situation.} 

It might be possible to limit $r$ by measurements of
the transverse Doppler effect; from the estimates for the so-called ``ether
wind'' $r c< 5 \times 10^{-10}$ holds \cite{Bogo07}.  

\section{Conclusion}
In this lecture, geometrizations of physics were mentioned some of which are
highly successful while others were not. It seems that all of them can be
related to exterior / interior symmetry groups (extension of Klein's ``Erlanger
Programm''?)\footnote{In this context, general relativity is seen as a gauge
  theory of some group.}. Why follow such an approach at all? There
are some advantages:\\
\begin{center} {\bf
- Geometrization helps to obtain new results in physics;\\
- Geometrization makes possible proofs of exact theorems in mathematical
 physics;\\
- Geometrization helps to ease (or even make possible) calculations in  physics.}\\ 
\end{center}
Perhaps, it is possible to distinguish two kinds of geometrization. The first 
type leads to a geometry forming only a {\em framework} for physical systems as
does special relativity. Nevertheless, the projective structure of Minkowski's
space-time can be related to the paths of free test particles; its
conformal structure to light and electromagnetic test signals. But the
second type has more structure. In it physical {\em fields} are
related with geometric objects (general relativity, Kaluza's
theory). In the mixed geometry of the 
Einstein-Schr\"odinger unified field theory there is too much structure to be 
useful for physics. This may also turn out to be the case with geometrization 
in the framework of Finsler geometry. This is yet as speculative as are 
geometrizations involving supersymmetry. \footnote{We have only
  mentioned but not dealt with areas in material physics in which
  Finsler geometry can be adapted to the structure of matter.} 

One central motive behind the urge for geometrization seems to be a 
wish for the {\em unification} of all fundamental interactions. Nevertheless,
we must insist that ``unification'' and ``geometrization'' are
separate concepts not necessarily forming a logical union.

Whether his philosophical conclusions about space and time are accepted, or 
not, the unification of the temporal and spatial aspects of physical reality
in space-time by mathematician Hermann Minkowski was a decisive step for all
later geometrizations. In this tribute to him I wanted to point out that, in
addition to his highly successful geometrization of space and time, he
also is indirectly connected - via Finsler geometry - to another type of 
geometrization. 

\bibliography{refs}

\begin{thebibliography}{999}

\bibitem{Riemann} Bernhard Riemann. {\it \"Uber die Hypothesen, welche der
  Geometrie zugrunde liegen.} Habilitationsvortrag in G\"ottingen am
  10. 6. 1854. (Nachdruck der Ausgabe von 1867 aus Band 13 der
  Abhandl. d. K\"oniglichen Gesellschaft d. Wissenschaften zu G\"ottingen:
  Dieterichsche Buchhandlung.) Darmstadt: Wissenschaftliche
  Buchgesellschaft (1959).
\bibitem{Schwarzschild00} Karl Schwarzschild. ``\"Uber das zul\"assige 
Kr\"ummungsma\ss{} des Raumes.'' {\it Vierteljahresschrift der astronomischen
  Gesellschaft} (Leipzig) {\bf 35}, 337-347 (1900). 
\bibitem{Ziegler} Renatus Ziegler. {\it Die Geschichte der
    geometrischen Mechanik im 19. Jahrhundert.} Wiesbaden: Franz Steiner (1985).
\bibitem{Minko1908} Hermann Minkowski. ``Raum und Zeit.'' Vortrag
  gehalten auf der 80. Naturforscher-Versammlung zu K\"oln am
  21. September 1908. Leipzig und Berlin: B. G. Teubner (1909).
\bibitem{Archi1920} R. C. Archibald. ``Time as a Fourth Dimension.''
  {\em Bulletin of the American Mathematical Society} {\bf 20},
  409-412 (1920).
\bibitem{Hinton1884}  Charles Howard Hinton. ``What is the fourth dimension?''
  {\it Scientific Romances} No. 1. London: Swan Sonnenschein \& Co. (1884).  
\bibitem{S1885} S. ``Four-Dimensional Space.'' {\it Nature} {\bf 31}, No. 804,
  March 26, 481. (1885)
\bibitem{Hinton1902}  Charles Howard Hinton. ``The recognition of the fourth
  dimension.'' {\it Bulletin of the Physical Society of Washington} {\bf 14},
  179 -203 (1902). Reprinted in R. v. Rucker {\it Speculations on the Fourth
    Dimension. Selected Writings of Charles H. Hinton.} New York: Dover 
Publications Inc. (1980)
\bibitem{Wells1895} Herbert George Wells. ``The Time Machine.'' London:
  William Heinemann (1895).
\bibitem{Greve1904} H. G. Wells. ``Die Zeitmaschine.'' Deutsch v. Felix
  P. Greve (Frederic Ph. Grove). Minden: Brund (1904). 
\bibitem{Palag1901} Melchior Pal\'agyi. {\it Neue Theorie des Raumes und der
    Zeit. Die Grundbegriffe einer Metageometrie.} Leipzig: Engelmann
  (1901). Auch in {\it Ausgew\"ahlte Werke}, Band III: ``Zur Weltmechanik. 
Beitr\"age zur Metaphysik der Physik'', pp. 1-33. Leipzig: Barth
(1925). [Nachdruck 1967 Wissenschaftliche Buchgesellschaft Darmstadt.]
\bibitem{Palag1914} Melchior Pal\'agyi. ``Zur Kritik der
  Relativit\"atstheorie,'' pp. 84-99 , in: {\it Ausgew\"ahlte Werke},
  Band III: ``Zur Weltmechanik. Beitr\"age zur Metaphysik der
  Physik.'' Leipzig: Barth (1925).
\bibitem{Poin05} Henri Poincar\'e.``Sur la dynamique de
  l'\'electron.''{\it Comptes rendus de l' Academie des Sciences,
    Paris} {\bf 140}, 1504-1508 (1905).
\bibitem{Goenner2008} Hubert Goenner. ``Max Plancks Beitr\"age zur speziellen
    Relativit\"atstheorie.'' submitted for publication (2008). 
\bibitem{Boyce1928} W: R. Boyce Gibson. ``The philosophy of Melchior Pal\'agyi. (I)
  Space-Time and the criticism of relativity.'' {\it Journal of Philosophical
  Studies} {\bf 3}, 15-28 (1928).
\bibitem{EinGross1914} A. Einstein and  M. Gro\ss{}mann. ``Entwurf einer
  verallgemeinerten Relativit\"atstheorie und einer Theorie der Gravitation.''
Leipzig: Teubner 1913. With additional remarks also in Zeitschr. f. Mathem. 
Physik,{\bf 62}, 225-261 (1914).
\bibitem{Einstein1915} A. Einstein. ``Die Relativit\"atstheorie.'' In: {\it Die
  Kultur der Gegenwart.} Dritter Teil, Dritte Abteilung, Band 1: Physik
(Red. E. Warburg), 703-713. Leipzig und Berlin: B. G. Teubner 1915.
\bibitem{Renn2007a} Michel Janssen, John D. Norton, J\"urgen Renn,
  Tilman Sauer, and John Stachel (eds.). {\it The genesis of general
    relativity.} Vols. 1\& 2: Einstein's Zurich Notebook. Boston
  Studies in the Philosophy of Science 250. Dordrecht: Springer (2007).
\bibitem{Renn2007b} J\"urgen Renn and Matthias Schemmel (eds.)
  Tilman Sauer, and John Stachel. {\it The genesis of general
    relativity.} Vols. 3\& 4: Gravitation in the twilight of classical 
physics. Boston Studies in the Philosophy of Science 250. Dordrecht: 
Springer (2007).
\bibitem{Busemann42} Herbert Busemann. {\it Metric Methods in Finsler Spaces
    and in the Foundations of Geometry.} Annals of Mathematics Studies,
  No. 8. Princeton: University Press 1942. Reprint 1965 New York: Kraus
  Reprint Co.
\bibitem{Thompson96} A. C. Thompson. {\it Minkowski
    Geometry}. Cambridge: University Press 1996.
\bibitem{Minkowski1910} Hermann Minkowski. {\it Geometrie der Zahlen.} Leipzig
  und Berlin: B. G. Teubner 1910.
\bibitem{Minkowski1911} Hermann Minkowski. {\it Volumen und Oberfl\"Ache.} In:
  {\it Gesammelte Abhandlungen von H. Minkowski}, Hrsg. D. Hilbert. Band II,
  p. 230-276. Leipzig und Berlin: B. G. Teubner 1911.
\bibitem{Gardner2006} Richard. R. Gardner. {\it Geometric Tomography.}
  Cambridge: University Press (2006).
\bibitem{Hilbert1910} David Hilbert. ``Hermann Minkowski.'' {\it
    Annalen der Mathematik} {\bf 68}, 445-471 (1910).
\bibitem{Busemann50} Herbert Busemann. ``The Geometry of Finsler Space.'' {\it Bulletin of the American Mathematical Society} {\bf 56}, 5-16 (1950).
\bibitem{Bragdon2005} Claude Fayette Bragdon. {\it A primer of higher
    space (the fourth dimension).} Cosimo classics: science. New York:
  Cosimo (2005).
\bibitem{Robbin2006} Tony Robbin. {\it Shadows of reality.\small The fourth
    dimension in relativity, cbism, and modern thought.}\normalsize New Haven
  \& London: Yale University Press (2006).
\bibitem{Papad2005} Athanase Papadopoulos. {\it Metric spaces,
    convexity and nonpositive curvature.} IRMA lectures in mathematics
  and theoretical physics ; 6. Z\"urich : Europ. Math. Soc. (2005).
\bibitem{Finsler1918} Paul Finsler. ``\"Uber Kurven und Fl\"achen in
  allgemeinen R\"aumen.'' Diss. Universit\"at G\"ottingen 1918; Nachdruck
  mit zus\"atzlichem Literaturverzeichnis. Basel: Birkh\"auser 1951.
\bibitem{Cartan1934} \'Elie Cartan. {\it Les Espaces de Finsler.} 
{\it Actualit\'es Scientifiques et Industrielles}, No. 79, Paris:
Hermann (1934). 
\bibitem{Matsu86} Makoto Matsumoto. {\it  Foundations of Finsler geometry
    and special Finsler spaces.} Kyoto: Kaiseisha Press (1986).
\bibitem{Matsu03} Makoto Matsumoto. ``Finsler Geometry in the
20th Century'', in vol. 1 of ``Handbook of Finsler Geometry'' (P.L. 
Antonelli, ed.), Kluwer Academic Publishers, Dordrecht/Boston/London 2003.
\bibitem{Hashi1986} Masao Hashiguchi. {\it Some topics on Finsler Geometry.}
Conferenze del Seminario di matematica dell' Universit\`aÂ  di Bari:
no. 210. Bari: Sem., Univ. (1986).
\bibitem{Mo2006} Mo, Xiaohuan. {\it An introduction to Finsler geometry}. Peking
University Series in Mathematics No. 1. Hackensack, NJ: World Scientific (2006).
\bibitem{Ciesz01} M. Cieszko. ``Fluid mechanics in Minkowski
  Space. Modeling of fluid motion in porous materials with anisotropic
  pore space structure.'' In: {\it IUTAM Symposium on Theoretical and
  Numerical Methods in Continuum Mechanics of Porous Materials}.
W. Ehlers ed., pp. 201-208. Dordrecht: Kluwer Academic Publishers 2001.

\bibitem{MikChic02} I.A. Miklashevich and A.V. Chigarev. ``Equation of
  the crack front with allowance for the metric properties of the
  material.'' {\it Russian Physics Journal} {\bf 45}, 1159-1164 (2002).

\bibitem{Bogo1973} George Yu. Bogoslovsky. ``On a special relativistic theory of
  anisotropic space-time.'' {Doklady Akad. Nauk SSSR} {\bf 213}, 1055-1058
  (1973). 
\bibitem{Bogo1977} G.Yu. Bogoslovsky. ``A special relativistic theory of the
  locally anisotropic space-time.'' Part I: The metric and group of
  motions of the anisotropic space of events.'' {\it Nuovo Cim.} {\bf
    40B}, 99-115(1977). Part II: ``Mechanics and electrodynamics in
  the anisotropic space.'' {\it Nuovo Cim.} 1977 {\bf 40B}, 116-134 (1977).
\bibitem{Bogo1994} G.Yu. Bogoslovsky. ``A viable model of locally anisotropic
  space-time and the Finslerian generalization of the relativity theory.''
  {\it Fortschritte der Physik} {\bf 42}, 143-193 (1994).
\bibitem{CohGlas06} A. G. Cohen and S. L. Glashow. ``Very special
  Relativity.'' {\it Physical Review Letters} {\bf 97}, 021601 (2006) 
\bibitem{Gibbons07} G. W. Gibbons, Joaquim Gomis and C. N. Pope. ``General
  very special relativity in Finsler geometry.'' archiv:0707.2174v2 20 Aug 2007.
\bibitem{GKZ1966} K. Greisen. ``End to the Cosmic-Ray Spectrum?'' {\it 
Physical Review Letters} {\bf 16}, 748-750 (1966).
\bibitem{Auger07} The Pierre Auger Collaboration (2007). ``Correlation
  of the Highest-Energy Cosmic Rays with Nearby Extragalactic
  Objects''. {\it Science} {\bf 318} (5852): 938-943 (2007). 
\bibitem{KostMew07} A. Kostelecky and M. Mewes, ``Lorentz-Violating
  Electrodynamics and the Cosmic Microwave Background.'' {\it
    Phys. Rev. Lett.} {\bf 99}, 011601 (2007). 
\bibitem{BluKostel2005} R. Bluhm, V. A. Kosteletzky et
  al. ``Spontaneous Lorentz Violation, Nambu-Goldstone Modes, and
  Gravity.'' {\it Physica Review} {\bf D 71}, 065008 (2005).
\bibitem{ColeGlash98} S. Coleman and S. L. Glashow ``High-energy tests
  of Lorentz invariance.'' {\it Physical Review} {\bf D 59}, 116008
  (1999); arXiv:hep-ph/9808446 (1998). 
\bibitem{CohGlash2006} A. G. Cohen and S. L. Glashow. ``Very special
  relativity''. arXiv:hep-ph/0601236v1 27. Jan. (2006).
\bibitem{Bogo07} George Yu. Bogoslovsky. ``Some physical displays of the space
  anisotropy relevant to the feasibility of its being detected at a
  laboratory.'' archiv:0706.2621v1 18 June 2007.
\end{thebibliography}

\pagebreak

\section{Glossary}

\begin{itemize}

\item Algebraic numbers

A complex number which is a root of a non-zero polynomial with rational
(or integer) coefficients (e.g., $\sqrt{3} $). An algebraic {\em
  integer} is a number which is a root of a non-zero polynomial with integer
coefficients (e.g., $1 \pm \sqrt{5}~; a + b i, a,b $
integers). Non-algebraic complex numbers are said to be transcendental 
(e.g., $\pi, e$).

\item Algebraic number field

A finite extension of the rational numbers ${\bf Q}$ is a ring of
algebraic integers ${\bf O}$ in an algebraic number field ${\bf
  K/O}$. The unique factorization of integers into prime numbers can
fail in ${\bf  K/O}$ (e.g., $6 = 2\cdot 3= (1+i\sqrt{5})\cdot (1-i\sqrt{5}))$.

\item Borel algebra

The Borel algebra (or Borel $\sigma$-algebra) on a topological space X is a 
$\sigma$-algebra of subsets of X associated with the topology of X. In the 
mathematical literature, there are at least two nonequivalent
definitions of this $\sigma$-algebra, either ``the minimal $\sigma$-algebra
containing the {\em open} sets'', or ``the minimal $\sigma$-algebra 
containing the {\em compact} sets''.

\item Convex set

A set $A \in E^n$ is convex if together with any two points $x,y;
~x\neq y \in A$ it contains the segment $[x,y]$, thus if $ (1-\lambda)x +
\lambda y \in A$ for $0 \leq \lambda \leq 1~.$

\item Convex hull~ $conv A$:

The intersection of all closed convex sets containing a given set $A$. Or: For
$A \in E^n$ the set of all convex combinations of any finitely many elements
of $A$.~ $conv(A + B) = conv A + conv B$. 

\item Homothetic

Sets $A, B$ are called homothetic if $A = \lambda B + d$ with $d \in
E^n, \lambda > 0.$

\item Mixed volume

Take a Minkowski sum of $r$ convex bodies: $\alpha_1 K_1 +\alpha_2 K_2
+ ...+ \alpha_r K_r ~,~ \alpha_i \geq 0~.$ The volume $\lambda(K)$ may
be written as a polynomial in the $\alpha_i$ with coefficients $V(K_1,
K_2,.. K_r)= V_{i_1 i_2....i_r}$ such that $\lambda(K) =
\Sigma_{i_{\mu} =1}^{r} V_{i_1
  i_2....i_r}\alpha_1\alpha_2...\alpha_r~. $ The coefficients $V_{i_1
  i_2....i_r}\alpha_1\alpha_2...\alpha_r$ may be taken as totally
symmetric in their indices; they are termed the {\em mixed volume}.

\item Length spaces

A metric space $X$ is a length space if for every $x,y$ in $X$
\begin{equation} |x-y| = \underset{\gamma}{inf}{~L(\gamma)} \nonumber \end{equation}
with L being the integral over the path $\gamma:[a,b] \rightarrow E^n$:
\begin{equation}L(\gamma)= \int_a^b \parallel \frac{d\gamma(t)}{dt}\parallel
  dt~.\nonumber \end{equation} The infimum is taken over the set of paths $\gamma$
  joining $x$ and $y$. 

\item Measure and volume

For a finite dimensional space $V$ there exists a single Hausdorff linear
topology. Thus the concept of Borel set (i.e., an element of a Borel
algebra) is intrinsic to $V$. A translation invariant measure on the
Borel $\sigma$-algebra is the so-called {\em Haar}-measure. This can
be used for the {\em volume} $\lambda (\cdot)$ in $V$.

A simpler way to introduce a volume is to assume that $V$
 is equipped with an auxiliary euclidean structure and that the volume
 $\lambda (\cdot)$ is the Lebesque measure induced by this
 structure. The particular choice of Haar-measure is immaterial. (A
 scalar multiple corresponds to a basis change for the euclidean
 structure.)

The volume of the d-dimensional euclidean unit ball is $ \frac{\pi^{\frac{d}{2}}}{\Gamma(\frac{d}{2}+1)}~.$

\item Orthogonality in normed spaces

If $V$ is a normed linear space and if $x,y \in V$, then $x$ is defined
to be orthogonal to $y$ if $\parallel x+\alpha y\parallel
\geq \parallel x \parallel$ for all $\alpha$ in $R$. 


\item Prime ideal

A prime ideal is a subset of a ring sharing important properties of a
prime number in the ring of integers. Any prime ideal of ${\bf Z}$ is of the
form $p{\bf Z}$, with $p$ a prime number. Ideals in ${\bf O}$ formed
with a prime number may no longer be a prime ideal, e.g., $2{\bf
  Z[i]}$, because $2{\bf Z[i]}= ((1+i)){\bf Z[i]})^2$. Fermat's
theorem says that for an odd prime number p \begin{center} $p{\bf
    Z[i]}$ is a prime ideal if $p \equiv 3~ (mod 4)$\\ $p{\bf
    Z[i]}$ is not a prime ideal if $p \equiv 1 (mod 4)$. \end{center}

Algebraic number theory generalizes this result to more general rings
of integers. 

\item Support function

Let $K \in E^n$ be a convex closed, non-empty body and $ u = (u_1,
u_2,...,u_n) \neq (0, 0,...., 0)$ a vector, and $x \in K $. Then the equation
of the support function can be written as $ sup\{ \Sigma x\cdot u |x \in K \} = 
h(K, u)$. $ \Sigma x\cdot u$ is the interior product in $E^n$. 

\item Support plane

Let $A \in E^n$ be a subset and $H \in E^n$ a hyperplane; let $H^+, H^-$ denote
the two closed half spaces bounded by $H$. We say ``$H$ supports $A$ at $x$'' if
$x \in A \cap H$ and either $A \in H^+$ or $A \in H^-$. $H$ is a {\em support 
plane} of $A$ or supports $A$ if $H$ supports $A$ at some point $x$ which is
necessarily a boundary point.

Let $A \in E^n$ be convex and closed. Then through each boundary point of $A$
there is a support (hyper)plane of A. If $A \neq 0$ is bounded, then to each 
vector u $u \in E^n \{ 0 \}$ there is a support plane to A with exterior
normal vector $u$. 

\end{itemize}

\end{document}